\documentstyle[multicol,epsf,aps]{revtex}

\begin{document}
\draft
\title{Fluctuation-dissipation ratio of the Heisenberg spin glass
}
\author{Hikaru Kawamura}
\address{Department of Earth and Space Science, Faculty of Science,
Osaka University, Toyonaka 560-0043,
Japan}
\date{\today}
\maketitle
\begin{abstract}

Fluctuation-dissipation (FD) relation of the three-dimensional
Heisenberg spin glass with weak random anisotropy is studied
by off-equilibrium Monte Carlo simulation.
Numerically determined FD ratio exhibits a ``one-step-like'' behavior,
the effective temperature of the spin-glass state being  about twice the
spin-glass
transition temperature, $T_{{\rm eff}}\simeq 2T_g$, irrespective of the
bath temperature.
The results are discussed in conjunction
with the recent experiment by H\'erisson and Ocio, and
with the chirality scenario of spin-glass transition.

\end{abstract}
\begin{multicols}{2}
\narrowtext

Off-equilibrium dynamics of spin glass (SG) has attracted much
recent interest\cite{review}. In equilibrium, there holds a
relation between the response and the correlation
known as the fluctuation-dissipation theorem (FDT).
In off-equilibrium, relaxation of physical
quantities of SG depends on its previous history. Most typically,
it depends not only on the observation time $t$ but also on
the waiting time $t_w$, {\it i.e.\/}, exhibits aging.
While in the short-time
quasi-equilibrium regime $t_0<<t<<t_w$ ($t_0$ is a microscopic
time scale), relaxation is still stationary and FDT
holds, it becomes
non-stationary and FDT is broken in the
long-time aging regime $t>>t_w$. In particular, the
breaking pattern of FDT, and its possible relation to the static quantity,
has been the subject of recent active studies.

A quantity playing a central role here
is the so-called fluctuation-dissipation ratio (FDR) $X$,
which may be defined by the relation,
\begin{equation}
R(t_1,t_2)=\frac{X(t_1,t_2)}{k_BT}\frac{\partial C(t_1,t_2)}{\partial t_1},
\end{equation}
where $R(t_1,t_2)$
is a response function measured at time $t_2$ to an impulse field
applied at time $t_1$, $C(t_1,t_2)$ is a two-time correlation function
in zero field
at times $t_1$ and $t_2$, and $T$ is the bath temperature.
One may regard $T/X\equiv T_{{\rm eff}}$ as an effective temperature.
In the case FDT holds, one has $X=1$ and $T_{{\rm eff}}=T$.

Via the study of certain mean-field
models, Cugliandolo and Kurchan showed that,
in the limit of infinite
time $t_1,t_2\rightarrow \infty$, the FDR $X$ depended
on the times $t_1$ and $t_2$ only
through the correlation function $C(t_1,t_2)$, {\it i.e.\/},
$X(t_1,t_2)=X(C(t_1,t_2))$ \cite{CK}.
It was further suggested
that $X(C)$ could be related to an appropriate
static quantity, {\it i.e.\/},
$X(C)$ is equivalent to the $x(q)$-function  describing the
replica-symmetry breaking (RSB)
pattern in the Parisi's scheme, which is related to the overlap distribution
function via
$P(q)=\frac{{\rm d}x(q)}{\rm{d}q}$\cite{CK}.
This conjecture was supported both by
numerical simulation\cite{Marinari98} and analytic work\cite{Franz}
(see, however,  also ref.\cite{Yoshino}).

Experimental studies on the FDR of SG have been hampered
for a long time by the difficulty
in performing high-precision measurements of correlations (noise measurements),
although some interesting preliminary results were reported\cite{CGKV}.
Recently, however, a remarkable experiment by H\'erisson and Ocio for an
insulating Heisenberg SG CdCr$_{1.7}$In$_{0.3}$S$_4$ has eventually
opened a door to experimental access to the FDR of SG\cite{HO}.
Some comparison was already made between these expeirmental results and
the numerical
results obtained by off-equilibrium simulation on certain SG models.

Meanwhile, it should be remembered that  all numerical results of the FDR
so far available are limited to the {\it Ising}-like SG models,
which represent
an extremely anisotropic limit
\cite{Marinari98,Yoshino,Barrat,Parisi99,Marinari00,Rieger}.
By contrast, the SG material studied
in ref.\cite{HO}
was more or less Heisenberg-like.
Recent experimental and numerical studies suggested that both equilibrium and
off-equilibrium properties might differ  significantly between the Ising
and Heisneberg SGs\cite{Kawa98,HukuKawa,KawaIma,Campbell,Vincent,Nordblad}.
Under such circumstances, in order
to make a direct link between theory and experiment,
it is clearly desirable to study
the Heisenberg-like SG where magnetic anisotropy is weak.
The purpose of the present Letter is to perform extensive
off-equilibrium simulations
on such a weakly anisotropic Heisenberg-like SG model, and numerically
investigate its FDR.
Interestingly,
the calculated FDR of the three-dimensional (3D) Heisenberg SG
exhibits a feature of the so-called one-step RSB.

I consider the weakly anisotropic classical
Heisenberg model 
whose Hamiltonian is given by
\begin{equation}
{\cal H}=-\sum_{<ij>} (J_{ij}{\mit\bf S}_i\cdot
{\mit\bf S}_j+D_{ij}^{\mu \nu }S_i^\mu S_j^\nu ) - h \sum_i S_i^z,
\end{equation}
where  ${\bf S}_i$ =$(S_i^x,S_i^y,S_i^z)$
is a three-component unit vector at the $i$-th site, and
the sum runs over all nearest-neighbor pairs of a 3D simple cubic lattice
with $N=L^3$ spins. $J_{ij}$ is the isotropic random exchange coupling
taking the value $J$ or $-J$ with equal probability ($\pm J$ or binary
distribution),  $D_{ij}^{\mu \nu}$
($\mu, \nu =x,y,z$) is the random exchange anisotropy which is assumed
to be symmetric ($D_{ij}^{\mu \nu}=D_{ij}^{\nu \mu}$) and
uniformly distributed between the interval [$-D$,$D$], and $h$ is a magnetic
field applied along $z$-direction. Aimed at clarifying the properties in the
limit of weak anisotropy,
I assume here a rather weak anisotropy of 
$D/J=0.01$. The SG transition
temperature of this model estimated from equilibrium Monte Carlo (MC)
simulation
is $T_g/J\simeq 0.21$\cite{HukuKawa2}.

Dynamical Monte Carlo
simulation is performed according to the standard single
spin-flip heat-bath method. At time zero,
I quench the system
from an infinite temperature to a working temperautre $T$ below $T_g$,
and wait in zero field
during the time $t_w$. In calculating the response,
a weak DC field of intensity  $h$ is applied, and
subsequent growth of the magnetization per spin,
[$<m_z(t+t_w)>$], is recorded at time $t+t_w$
($<\cdots >$ and [$\cdots $] are thermal
average and average over quenched randomness).
In calculating correlation, I measure
the two-time correlation function of magnetization
at times $t_w$ and $t+t_w$,
$C(t;t_w)=[<\vec m_z(t_w)\cdot \vec m_z(t+t_w)>]$,
always staying in zero field in this case.

The integrated FD relation eq.(2)
yields the zero-field cooled (ZFC) susceptibility
$\chi \equiv [<m_z(t_w+t)>]/h$ as,
\begin{equation}
\chi(t;t_w)=\frac{1}{3k_BT}\int ^1_{C(t;t_w)}X(C'){\rm d}C',
\end{equation}
where 
the factor $\frac{1}{3}$
takes care of the number of components of Heisenberg spins.

The lattice size mainly studied is $L=32$ with periodic
boundary conditions.
I generate total of $10^6$
Monte Carlo steps per spin (MCS)
in each run, with $t_w=2.5\times 10^4$ or $10^5$  MCS.
Sample average is taken over 40-160 independent bond realizations.

The calculated spin  autocorrelation functions
are shown in Fig.1 as a function of observation time scaled
by the waiting time, $t/t_w$, for several
temperatures below $T_g$.
The decay of the autocorrelation at lower temperatures
occurs in two steps. Near the end of the
quasi-equilibrium regime,  the decay curves tend to level off exhibiting
a plateau-like structure, which is expected
to correspond to the static
Edwards-Anderson order parameter $q_{{\rm EA}}$\cite{Kawa98}.
Furthermore, 
at lower temperatures $T/J=0.10$ and 0.05
the two curves for
different $t_w$ cross around $t/t_w\sim 1$, and at $t>t_w$,
the data for larger $t_w$ lie {\it above\/} the one
for smaller $t_w$ (superaging).
Such a superaging behavior means
that the aging is apparently
more enhanced than the one expected from the naive
$t/t_w$-scaling. The observed superaging of spin correlations is
similar to the one previously observed in the Heisenberg SG with the
Gaussian coupling\cite{Kawa98}.
At higher temperatures $T/J=0.15$ and 0.20, by contrast,
the curve for longer
$t_w$ lies slightly {\it below\/} the one
for shorter $t_w$ (subaging), or nearly comes on top of it 
for longer $t$ (near full aging). 
To examine the robustness of the observed aging 
behavior against the system size and the waiting time, I show in the inset 
the autocorrelation for the smaller size $L=16$ for the temperatures
$T/J=0.15$ and 0.05, with $t_w=5\times 10^3$ and $2.5\times 10^4$.
The superaging/subaging
tendency observed for $L=32$ can also be seen here.

Hence, whether the spin
correlation exhibits superaging or subaging apparently
depends on the bath temperature,
superaging at lower temperatures and subaging (or near full aging)
at higher temperatures. In the present case, the borderline temperature
is around $T/J\simeq 0.15$, corresponding to $\simeq 0.7T_g$.
In the recent noise experiment by H\'erisson and Ocio, subaging
behavior is observed at $T\simeq 0.8T_g$\cite{HO}.
It might be interesting to see experimentally
whether the superaging behavior as observed here ever
arises at lower temperatures.
In real experiment, however, care needs to be taken to an inevitable
cooling-time effect which might give some bias
toward apparent subaging\cite{Berthier,Orbach}.

\begin{figure}[h]
\epsfxsize=\columnwidth\epsfbox{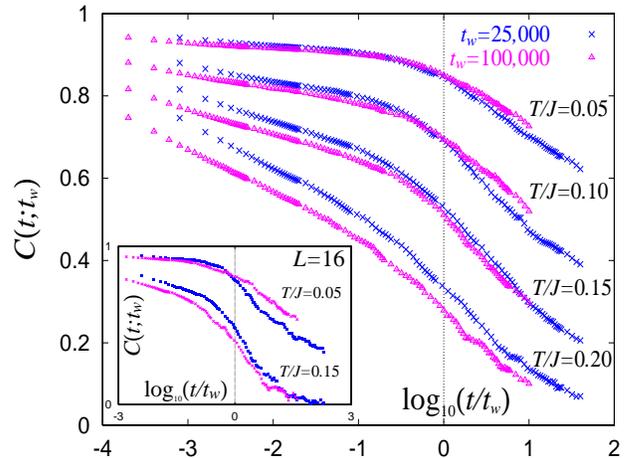}
\caption{
Decay of the spin autocorrelation function of the weakly
anisotropic 3D Heisenberg spin glass with $D/J=0.01$, plotted versus
$\log _{10}(t/t_w)$ at several temperature below $T_g$.
The lattice size is $L=32$, averaged over 80 ($T/J=0.20$), 
160 ($T/J=0.15$) and 40 ($T/J=0.10$) samples. Inset:  
The autocorrelation decay for the smaller system of
size $L=16$ averaged over 40 samples. The waiting times are 
$t_w=5\times 10^3$ (blue) and $2.5\times 10^4$ (pink) here.
}
\end{figure}

In the lower panel of Fig.2, I show the growth of the ZFC susceptibility
$T\chi (t;t_w)$ as a function of
$t/t_w$ at several temperatures.
I applied a very week field of intensity
$h/J=0.01$.
A word of caution is in order here. Generally, applied fields should be
sufficiently weak so that one is in the linear response regime.
In the previous simulations on the Ising SG, the
field intensity of $h/J=0.1$ turned
out to be weak enough\cite{Marinari98,Yoshino,Parisi99}.
In the case of
Heisenberg SG, however, still weaker field appears
to be necessary to guarantee the linearity of the response.
This is illustrated in the upper panel of Fig.2 where I show the
$t$-dependence of the ZFC susceptibility at $T/J=0.15$
for three different field intensities,
$h/J=0.05,0.02$ and $0.01$. Some deviation is discernible
between these curves at longer $t$, suggesting that the
requirement of the linearity is rather severe in the Heisenberg
case. In the following, I show the response
data for the weakest field studied, $h/J=0.01$,
where the linearity seems almost satisfied in the investigated time range.

Indeed, such a strong nonlinearity of the response appears to be even more
enhanced in the isotropic limit $D=0$. I also
made a preliminary simulation on the
fully isotropic Heisenberg SG with $D=0$ at a temperature $T/J=0.15$
to study its response, with varying the
field intensity $H/J=0.05,\ 0.02,\ 0.01$ and 0.005.
While, for a field of moderate intensity $H/J=0.05$,
the response $\chi(t)$ of the isotropic system $D=0$
is almost common with the one of the weakly
anisotropic system $D/J=0.01$, on decreasing the field intesnity, $\chi(t)$
of the isotropic system constantly gets larger at longer observation time
beyond the corresponding
value for $D/J=0.01$, without showing any tendency of saturation
up to $H/J=0.005$. This strongly suggests that the field scale associated
with the linear response is closely correlated with the strength of
the mangetic anisotropy $D$, although its detailed properties are
yet to be clarified.


\begin{figure}[h]
\epsfxsize=\columnwidth\epsfbox{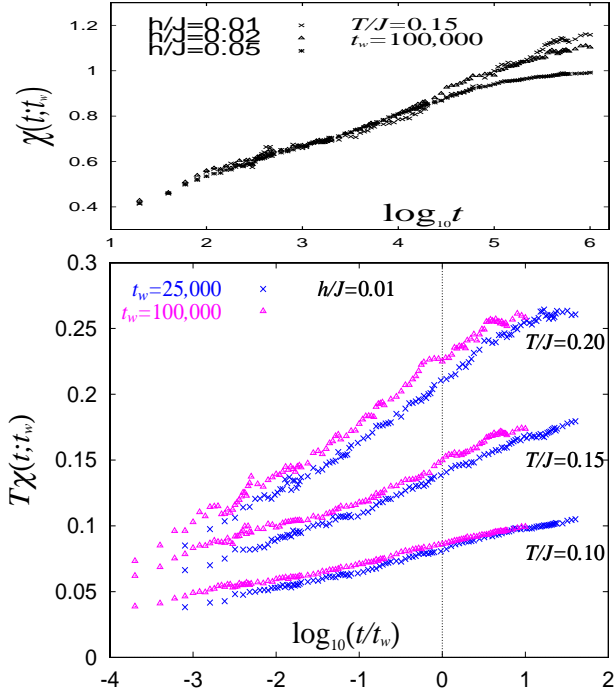}
\caption{
Growth of the ZFC  susceptibility $\chi(t;t_w)$ of
the weakly anisotropic 3D Heisenberg SG with $D/J=0.01$.
The lattice size is  $L=32$,  averaged over 80 ($T/J=0.20$), 
160 ($T/J=0.15$) and 40 ($T/J=0.10$) samples.
Upper panel: $\chi(t;t_w)$ plotted versus
the observation time $t$ for several applied field intensities.
The temperature is $T/J=0.15$, with the waiting time $t_w=10^5$ MCS.
Lower panel: $T\chi(t;t_w)$ plotted versus
$\log _{10}(t/t_w)$ at several temperatures below $T_g$, with
$t_w=2.5\times 10^4$ and $10^5$ MCS.
}
\end{figure}

As can be seen from the lower panel of Fig.2,
the curve for longer
$t_w$ tends to lie slightly above the one
for shorter $t_w$ (subaging), whereas  for $t>>t_w$
the two curves for different $t_w$ merge asymptoticaly
(full aging).
The superaging behavior, which has
been seen in the correlation, is not clearly visible here in the reponse.
Experimentally, response of SG usually exhibits a weak
subaging\cite{review,HO} or
near full aging\cite{Orbach}, which
seems consistent with the present result.

In Fig.3(a), I show  a parametric plot $T\chi(t;t_w)$ versus $C(t;t_w)$
for several temperatures. The straight line in the figure is the FDT line,
$T\chi=(1-C)/3$.
In the short-time quasi-equilibrium regime, the data lie on the FDT line,
and deviate from the line in the long-time aging regime. In the
aging regime, the data for longer $t_w$ tend to lie above the data for
shorter $t_w$, slightly, but systematically. Such a systematic drift of
the data with varying $t_w$ is a common aspect observed
both in experiment\cite{HO}
and in simulations on the Ising SG
\cite{Marinari98,Yoshino,Barrat,Parisi99,Marinari00,Rieger}.

In Fig.3(b), the data for longer $t_w$ are replotted in the form of
$\chi(t;t_w)$ versus $C(t;t_w)$. One sees that the data in the aging reime
at various
temperatures lie on top of each other, indicating that $\chi$ is
temperature independent in the aging regime, apparently
satisfying Parisi-Toulouse hypothesis\cite{Marinari98,Marinari00}.

\begin{figure}[h]
\epsfxsize=\columnwidth\epsfbox{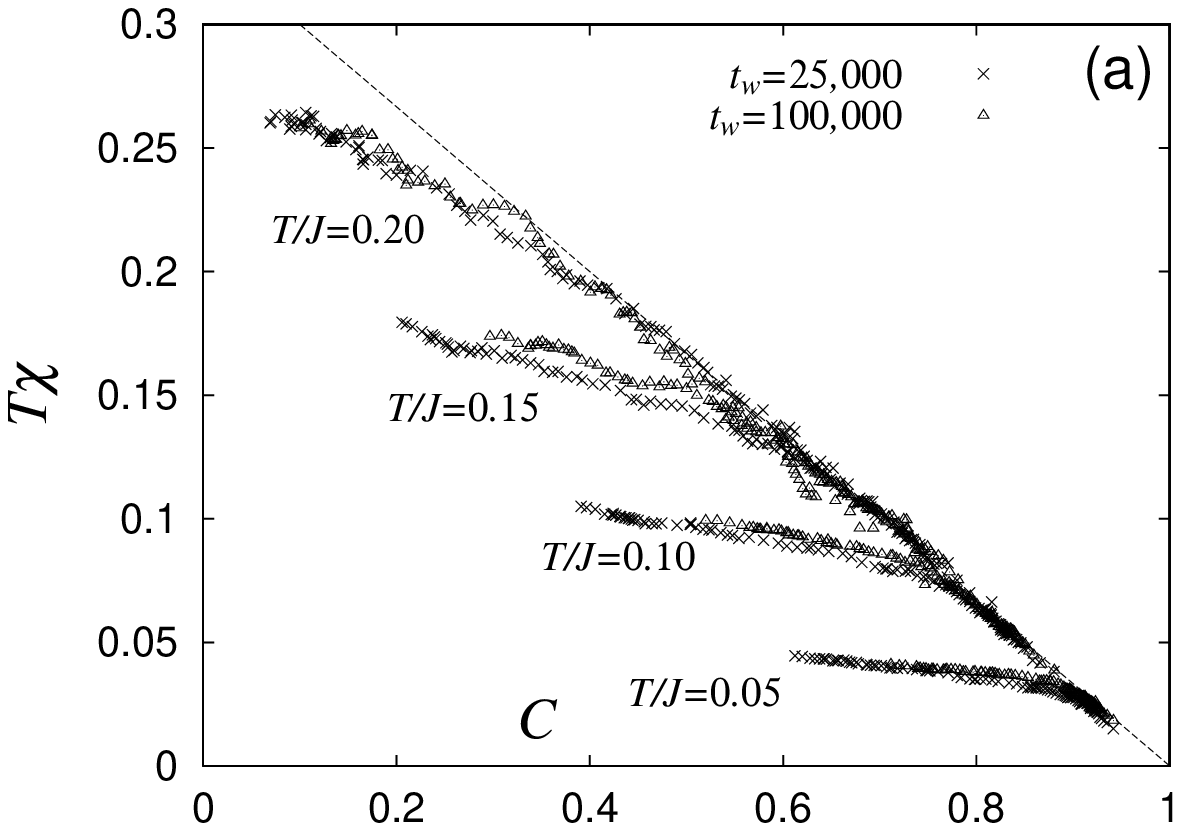}
\epsfxsize=\columnwidth\epsfbox{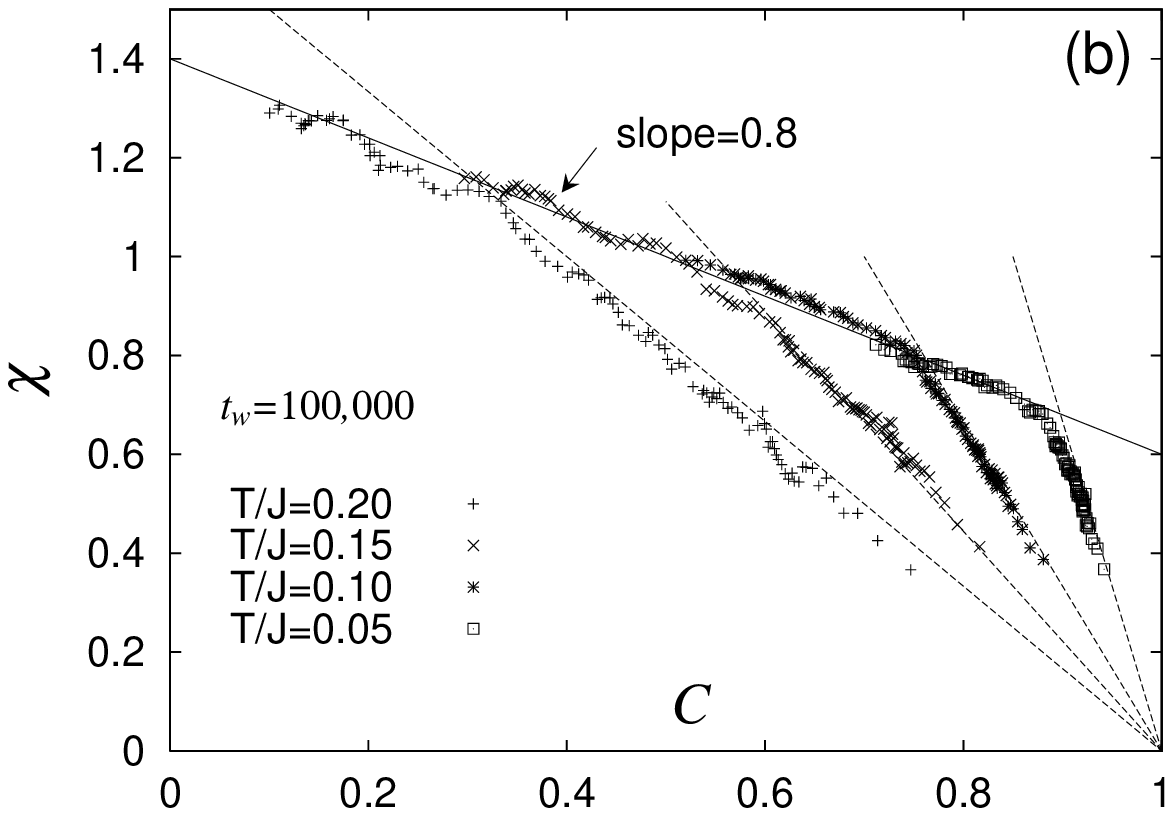}
\caption{ $T\chi$ versus $C$ plot (a), and $\chi$ versus $C$  plot (b),
of the weakly anisotropic
3D Heisenberg SG with $D/J=0.01$ at several
temperatures below $T_g$.
The applied field intensity is  $h/J=0.01$.
The lattice size is  $L=32$, averaged over 80 ($T/J=0.20$), 
160 ($T/J=0.15$) and 40 ($T/J=0.10,\ 0.05$) samples. 
In (b), $t_w$ is  $10^5$ MCS.
The broken lines represent the FDT lines. The solid line in (b)
is the straight-line fit to the data in the aging regime,
its slope being equal to 0.8.
}
\end{figure}

Remarkable in Fig.3
is the strikingly linearity of the plot
in the aging regime, although some curvature is discernible in
the part where the data begin to deviate from the FDT line.
In particular, the data in the aging regime
appear to come to the $C=0$ axis {\it with
a nonzero tangent\/}. Such a  behavior seems distinct with
the behavior of the Ising SG
where the data exhibit some more curvature, and appear to come to the $C=0$
axis almost horizontally, or at least be consistent with a vanishing
tangent at $C=0$\cite{Marinari98,Parisi99,Marinari00}.
By making a linear fit of the data in the aging regime in Fig.3(b),
I estimate the slope to be about 0.8,
which corresponds to the effective temperature
$T_{{\rm eff}}\simeq 0.42J\simeq 2T_g$,
irrespective of the bath temperature $T(<T_g\simeq 0.21J)$.

Such a one-step-like behavior of the FDR,
though familiar in structural glasses,
may sound surprising in SG. However, this observation
finds a natural explanation in the recently developed chirality scenario of
SG transition, which ascribes the true order parameter of the SG order
to the {\it chirality\/}\cite{Kawa98,Kawa92}.
In this scenario, chirality exhibits a one-step-like
RSB in the ordered state (chiral-glass state)\cite{HukuKawa},
which is manifested in the
SG order in real Heisenberg-like SG via the weak random magnetic
anisotropy. Thus, the chirality theory predicts the occurrence of
a one-step-like RSB in the SG ordered state of real Heisenberg-like SG,
which, in its off-equilibrium dynamics, might well be reflected
in a one-step-like behavior of the FDR.
In particular, a nonzero tangent is expected at $C=0$ for the FDR,
which corresponds
to the delta-function peak in the static
overalp distribution function $P(q)$
at $q=0$\cite{HukuKawa}.
The present numerical observation seems fully consistent with such an
expectation from the chirality scenario.

Now, I wish to compare the present numerical result of the FDR
with the recent experiments. The one-step-like linear behavior
observed in the present simulation seems at least not inconsistent
with the experimental data of refs.\cite{CGKV,HO},
although the data of ref.\cite{HO}
appear to exhibit some curvature showing a tendency to level off
toward $C=0$.
It would be interesting to experimentally explore further
deep into the aging regime (smaller $C$ regime), to see whether a
finite tangent persists or not at $C\rightarrow 0$. Meanwhile,
the experimentally determined
effective temperautre of the SG state of
CdCr$_{1.7}$In$_{0.3}$S$_4$ was
$T_{{\rm eff}}\simeq 30K\simeq 1.9T_g$ ($T_g\simeq$ 16.2K)
at the measuring temperature $T=13.3K$\cite{HO},
which is in good agreement with the
present estimate of $T_{{\rm eff}}$ given above. 

It should be noticed that CdCr$_{1.7}$In$_{0.3}$S$_4$,
though basically a Heisenberg-like magnet, might
possess a bit stronger anisotropy than the very weak anisotropy assumed
here ($D/J=0.01$). Such Ising-like tendency of 
CdCr$_{1.7}$In$_{0.3}$S$_4$ might be a possible cause of the curvature
observed in the $\chi$-$C$ plot of ref.\cite{HO}.
It would then be highly interesting,  both
experimentally and numerically, to investigate $T_{{\rm eff}}$ 
at various bath temperatures for 
various systems with varying degrees of anisotropy.

In summary, I studied the fluctuation-dissipation ratio of the
3D Heisneberg SG with weak random anisotropy by
off-equlibrium MC simulation. 
The calculated FDR exhibits a one-step-like behavior
with the effective temperature of the SG state
$T_{{\rm eff}}\simeq 2T_g$ irrespective
of the bath temperature. 
The observed one-step-like behavior of the FDR is
interpreted in terms of the recently developed chirality scenario
of SG transition.

The author is thankful to  M.Ocio, E.Vincent
and H.Yoshino for useful discussion.
The numerical calculation was performed on the Hitachi SR8000 at the
supercomputer center, ISSP, University of Tokyo.

\end{multicols}

\end{document}